\documentclass[12pt,titlepage,a4paper]{myarticle}

\usepackage[T1]{fontenc} 
\RequirePackage[colorlinks=true
,urlcolor=blue
,anchorcolor=blue
,citecolor=blue
,filecolor=blue
,linkcolor=blue
,menucolor=blue
,linktocpage=true
,pdfproducer=medialab
,pdfa=true
]{hyperref}
\usepackage{amsmath, mathrsfs, amsfonts, amssymb, amsthm, mathtools, graphicx, color, ucs, xparse, tikz, lmodern, physics, varioref, cleveref, breqn, tensor, cite}
\usepackage{hyphenat}

\crefname{pluralequation}{eqs.}{eqs.}
\Crefname{pluralequation}{Eqs.}{Eqs.}
\crefformat{pluralequation}{eqs.~(#2#1#3)}
\Crefformat{pluralequation}{Eqs.~(#2#1#3)}

\renewcommand{\d}{\partial}

\newcommand{\overbar}[1]{\mkern 2mu\overline{\mkern-2mu#1\mkern-2mu}\mkern 2mu}

\newcommand\mperiod[1][\rlap]{#1{\ .}}	
\newcommand\mcomma[1][\rlap]{#1{\ ,}}

\crefname{table}{table}{tables}
\Crefname{table}{Table}{Tables}
\crefname{figure}{figure}{figures}
\Crefname{figure}{Figure}{Figures}

\newenvironment{eqaed}
    {\begin{equation}
    \begin{aligned}
    }
    { 
    \end{aligned}
    \end{equation}
    \ignorespacesafterend
    }

\begin{document} 

\title{Fake supersymmetry with tadpole potentials}
\author{Salvatore Raucci
     \oneaddress{
     Scuola Normale Superiore and INFN \\ Piazza dei Cavalieri 7, 56126, Pisa, Italy \\
     {~}\\
      \email{salvatore.raucci@sns.it}
}}

\Abstract{We study tadpole potentials of non-supersymmetric strings, resorting to a first-order formalism known in the literature as fake supersymmetry. We present a detailed analysis for vacua with only gravity and the dilaton, displaying the obstructions that forbid the simplest inclusion of form fluxes. Our focus is on codimension-one vacua, for which we propose a definition of energy that might be suitable for stability arguments. Our findings point to the central role of boundary conditions when supersymmetry is absent or broken.}

\maketitle
\setcounter{page}{1}

\tableofcontents

\section{Introduction}

Stability is a subtle issue in a theory of gravity, ultimately because there is no meaningful concept of a state of lowest energy. String theory, while offering a consistent framework for a quantum theory of gravity, exhibits a variety of mechanisms that can generate unstable behavior. In fact, tachyons serve as instability probes in string perturbation theory, but we know that they are only part of the story. For instance, the open string spectrum of a system comprising a brane and an anti-brane, which are subject to a mutual attraction, is free of tachyons if their mutual separation is large enough. More catastrophic decays are possible, through bubble nucleation~\cite{Coleman:1980aw} or via generalizations of Witten's bubble of nothing~\cite{Witten:1981gj}.

We currently lack a worldsheet characterization of fully stable vacua, so that the spacetime picture is the main tool that can guide our investigations. Since one expects that in a quantum theory any allowed physical process will eventually happen, the stability analysis translates into the search for possible obstructions. 
In the low-energy quantum field theory regime, this calls for vacuum-protecting symmetries and, in fact, supersymmetry is usually the main culprit for the settings that we consider reliable.
Then, it is not surprising that severe backreactions typically accompany the breaking of supersymmetry. This is already evident in ten dimensions, where dilaton tadpoles emerge in tachyon-free string theories without supersymmetry: although these theories are originally formulated in flat space, this is not a true vacuum for them.

Understanding the most general types of obstructions remains an open question. We know, for instance, that even supersymmetry does not suffice to guarantee the stability of string vacua, and topological charges\footnote{In our language, the spin structure, which obstructs the bubble of nothing of~\cite{Witten:1981gj}, is an example of a topological charge.} seem in general a necessary ingredient~\cite{GarciaEtxebarria:2020xsr}.
In this work, we shall focus on the local conditions for stability. Leaving global and topological considerations for future work, we shall attempt to identify a replacement for supersymmetry when it is broken or absent. 
Our ultimate aim would be to attain a better understanding of string models with tadpole potentials, where the mechanism of~\cite{Fischler:1986ci, Fischler:1986tb} should resolve the effect of the runaway potential through deformations of the background. See for instance~\cite{Dudas:2004nd, Kitazawa:2008hv} for a string theory discussion regarding the models of interest in this paper.

Tadpole potentials are a generic hallmark in the absence of Fermi-Bose degeneracy, since the corrections to the vacuum energy expected from quantum field theory are dressed with exponentials of the dilaton. Our focus will be on the three ten-dimensional tachyon-free models without supersymmetry: the heterotic $\text{SO}(16)\times\text{SO}(16)$ string~\cite{AlvarezGaume:1986jb, Dixon:1986iz}, the U(32) orientifold~\cite{Sagnotti:1987tw, Pradisi:1988xd, Horava:1989vt, Horava:1989ga, Bianchi:1990yu, Bianchi:1990tb, Bianchi:1991eu, Sagnotti:1992qw} model~\cite{Sagnotti:1995ga, Sagnotti:1996qj}, and Sugimoto's USp(32) orientifold~\cite{Sugimoto:1999tx}. However, our approach could be extended to more general settings.

The feature that guides our analysis is that supersymmetric vacua can be identified by solving first-order equations determined by supersymmetry transformations, since these imply the equations of motion~\cite{Lust:2004ig, Gauntlett:2005ww, Lust:2008zd}, up to subtleties on the time-space components of the metric equations.
This property is also the rationale behind the geometrical structures that enter string compactifications and the spinorial definitions of energies~\cite{Witten:1981mf, Nester:1981bjx} that control the dynamical stability. 
The method that we discuss in this paper is known in the literature as fake supersymmetry~\cite{Boucher:1984yx, Townsend:1984iu, Skenderis:1999mm, Freedman:2003ax}. Our strategy, inspired by~\cite{Giri:2021eob}, is to include dilaton tadpoles of non-supersymmetric strings in a set of first-order equations that imply the equations of motion.
This procedure puts the dilaton potential on the same footing as the lowest-order terms in the equations of motion and is most suitable for intrinsically quantum vacua, using the terminology of~\cite{Baykara:2022cwj}.

In this paper, we shall be mainly interested in vacuum solutions for the non-supersymmetric strings in the absence of fluxes and charged sources. The reasons are technical, as we shall explain in \cref{sec:RR}, but we are also motivated by the perturbative stability~\cite{Basile:2018irz} of the Dudas-Mourad vacuum~\cite{Dudas:2000ff}, whose ultimate fate remains an open question.

In \cref{sec:fake_susy}, we introduce the fake supersymmetry formalism and apply it to the dilaton-gravity system that is relevant for the non-supersymmetric models of interest, commenting on the difficulties that one faces when R-R forms are present. Then, in \cref{sec:vacuum_solutions}, we recover the Dudas-Mourad solution in this new language, and in \cref{sec:energy} we define a Witten-Nester two-form, following~\cite{Giri:2021eob}, which could play the role of an energy for the codimension-one vacua. We comment on this interpretation in \cref{sec:stability_DM}.

\section{Fake supersymmetry}\label{sec:fake_susy}

In a supergravity theory, setting to zero the supersymmetry variations of the Fermi fields is the most convenient way to explore the supersymmetric landscape. As shown in~\cite{Lust:2004ig, Gauntlett:2005ww, Lust:2008zd}, if the Bianchi identities also hold, the equations of motion are automatically satisfied.\footnote{Ten-dimensional Majorana-Weyl spinors have a non-empty annihilator, and therefore some of the equations of motion are not implied by supersymmetry. However, when the metric ansatz is sufficiently symmetric, for instance in compactifications with a maximally symmetric external spacetime, the missing equations can be actually deduced from the others. We shall leave aside this technical discussion.}
In the main portion of this paper, we shall not include form fluxes and gauge fields, so that in a supersymmetric setup one would only impose the vanishing of the gravitino and dilatino variations, $\delta\psi_M=0$ and $\delta\lambda=0$.
In this simple case, with only the metric field and the dilaton, these schematically read
\begin{eqaed}
    \delta  \psi_M & = D_{M}^{\mbox{\scriptsize susy}}  \varepsilon=\nabla_M  \varepsilon=0\mcomma\\
    \delta  \lambda & =\mathcal{O}^{\mbox{\scriptsize susy}}  \varepsilon= d\phi  \varepsilon=0\mperiod
\end{eqaed}

The aim of this paper is to modify the two first-order differential operators in such a way as to include dilaton potentials in the equations of motion. In this basic setting with gravity and the dilaton, this strategy has already been pursued in the literature, and goes under the name of fake supersymmetry~\cite{Freedman:2003ax}.
It is essential to understand that this procedure has no physical implications because the new susy-like equations do not correspond to any symmetry of the system. The modified operators $D_M$ and $\mathcal{O}$ that we define are merely a technical tool, whose purpose is to recover the correct equations of motion.
This being said, we thus start with the two conditions
\begin{eqaed}\label[pluralequation]{eq:fake_susy_general}
    D_{M}  \varepsilon & =0 \mcomma \\
    \mathcal{O} \varepsilon & = 0 \mcomma
\end{eqaed}
and deduce the form of the operators by demanding that they imply the equations of motion in the presence of a non-trivial scalar potential.

In general, we expect that \cref{eq:fake_susy_general} yield only a subset of the solutions to the equations of motion, but for this subset they allow a systematic way to engineer and analyze vacua, as we shall explain. We shall comment further on these limitations in \cref{sec:some_comments}.

\subsection{Tadpole potential in the string frame}\label{sec:tadpole_string_frame}

Let us consider the following string-frame action in ten dimensions
\begin{eqaed}\label{eq:action_gravity_dilaton}
    \int d^{10}x \ \sqrt{-g} \left[e^{-2\phi}\left(R+4(\d\phi)^2\right)-V(\phi)\right]\mperiod
\end{eqaed}
With appropriate choices for the potential, this serves as the low-energy effective action of the ten-dimensional non-supersymmetric strings, see~\cite{Mourad:2017rrl} for a review.
The equations of motion read
\begin{eqaed}\label[pluralequation]{eq:equations_of_motion_gravity_dilaton}
    R_{MN}+2\nabla_M\nabla_N\phi+\frac{1}{2}e^{2\phi}\left(V+\frac{1}{2}V'\right) g_{MN} & = 0\mcomma \\
    R+4\nabla^2\phi-4(\d\phi)^2+\frac{1}{2}e^{2\phi} V'&=0\mcomma
\end{eqaed}
where $V'=\d_\phi V(\phi)$. We define the two operators that should replace the gravitino and dilatino variations as
\begin{eqaed}\label[pluralequation]{eq:fake_susy_eq_variations}
    D_{M} \varepsilon & =\left(\nabla_M + \mathcal{W}(\phi) \Gamma_M\right)\varepsilon\mcomma \\
    \mathcal{O} \varepsilon&= \left(d\phi+g(\phi)\right)\varepsilon\mperiod
\end{eqaed}
$\mathcal{W}(\phi)$ is known as a fake superpotential because it enters the first equation as a supersymmetric superpotential would.
Note that it is not even necessary for the model to have a gravitino or a dilatino, as happens for instance in type 0'B, because, as we already stressed, the above equations are nothing more than a formal tool.
There is also no restriction on the spinor $\varepsilon$. We could consider an arbitrary number of spinors, adding an index to \cref{eq:fake_susy_eq_variations}. For the moment, let us consider the simplest possibility of a single spinor $\varepsilon$, with the possible restriction to a Majorana-Weyl spinor, if needed.

In order to recover the first of \cref{eq:equations_of_motion_gravity_dilaton}, we employ an appropriate combination of the above operators, following the supersymmetric procedure reviewed for instance in~\cite{Tomasiello:2022dwe}:
\begin{eqaed}\label{eq:string_gravity_eom}
    0&=2\bigg[\Gamma^M\comm{D_M}{D_N}+\comm{D_N}{\mathcal{O}}+(\mathcal{W}'-2\mathcal{W})\Gamma_N\mathcal{O}\bigg]\varepsilon=\\
    &=\bigg[\Gamma^M (R_{MN}+2\nabla_M\nabla_N\phi)+(36\mathcal{W}^2+2 \mathcal{W}'g - 4 \mathcal{W} g)\Gamma_N + \\ & +(2g'-4\mathcal{W}-16\mathcal{W}')\nabla_N\phi\bigg]\varepsilon\mperiod
\end{eqaed}
Similarly, for the second of \cref{eq:equations_of_motion_gravity_dilaton},
\begin{eqaed}\label{eq:string_dilaton_eom}
    0&=\bigg[(D-\mathcal{O})^2 \! - \! (\nabla_M-2\d_M\phi)D^M+(g'-2g-9\mathcal{W}'+18\mathcal{W})\mathcal{O}-(19\mathcal{W}-2g) D \bigg]\varepsilon= \\
    & = \left[-\frac{1}{4}\left(R+4\nabla^2\phi-4(\d\phi)^2\right)+(-9\mathcal{W}'g + g g'+18 \mathcal{W} g - g^2 -90 \mathcal{W}^2)\right]\varepsilon\mcomma
\end{eqaed}
where $D=\Gamma^M D_M$.
We then retrieve \cref{eq:equations_of_motion_gravity_dilaton} from \cref{eq:string_gravity_eom} and \cref{eq:string_dilaton_eom} provided 
\begin{eqaed}\label[pluralequation]{eq:the_three_conditions}
    36\mathcal{W}^2+2\mathcal{W}'g-4\mathcal{W} g & = \frac{1}{2}e^{2\phi}\left(V+\frac{1}{2} V'\right)\mcomma \\
    g'-2\mathcal{W}-8\mathcal{W}' & = 0\mcomma\\
    -9\mathcal{W}'g+ g'g +18\mathcal{W} g -g^2-90\mathcal{W}^2 & = -\frac{1}{8}e^{2\phi} V'\mperiod
\end{eqaed}
One of the three conditions is redundant, and in terms of 
\begin{eqaed}
    h(\phi)\equiv g(\phi)-6\mathcal{W}(\phi)\mcomma \qquad k(\phi)=g(\phi)-12 \mathcal{W}(\phi)\mcomma
\end{eqaed}
the two independent equations take the form
\begin{eqaed}\label[pluralequation]{eq:string_frame_two_conditions}
    hk&=-\frac{1}{4}e^{2\phi}V \mcomma\\
    h-k & = 2h'+k'\mperiod
\end{eqaed}
Let us concentrate on the potentials that arise in non-supersymmetric strings,
\begin{eqaed}\label{eq:tadpole_potential}
    V(\phi)=T e^{\alpha\phi}\mcomma
\end{eqaed}
with $\alpha=0$ and $-1$ for the heterotic and the orientifold models respectively. Note that the former value of $\alpha$ is appropriate for more general models in which there is a non-zero one-loop vacuum energy contribution in the string frame.
There is a sign ambiguity in \cref{eq:string_frame_two_conditions}, namely $(h,k)\to (-h,-k)$, and for this reason we define additional functions $A(\phi)$, $B(\phi)$ as
\begin{eqaed}
    hk=-A\mcomma \qquad \frac{k}{h}=-2B\mcomma
\end{eqaed}
so that $A$ and $B$ are always positive for the potentials of \cref{eq:tadpole_potential}. Then, \cref{eq:string_frame_two_conditions} read
\begin{eqaed}\label[pluralequation]{eq:A_and_B_eq}
    A&=\frac{T}{4}e^{(2+\alpha)\phi}\mcomma\\
    \frac{B'}{B}+B' & = (1+\alpha)-(4+\alpha)B\mperiod
\end{eqaed}
Depending on the value of $\alpha$, there are different types of solutions to the second of \cref{eq:A_and_B_eq}:
\begin{itemize}
    \item For $\alpha<-4$ and $\alpha> -1$, the $B$ equation has a constant solution
    \begin{eqaed}\label{eq:constant_B}
        B=\frac{1+\alpha}{4+\alpha}\mperiod
    \end{eqaed}
    \item For $\alpha=-4$, one can rewrite the second of \cref{eq:A_and_B_eq} in terms of the variable $e^{-3\phi}$, discovering the differential equation that defines the Lambert $W$ function,\footnote{The algebraic definition of the Lambert function is $W(x)$ such that $W(x)e^{W(x)}=x$.} hence 
    \begin{eqaed}\label{eq:B_for_alpha_-4}
        B=W(e^{-3(\phi-c)})\mcomma
    \end{eqaed}
    with $c$ a constant.
    \item For $\alpha=-1$, similar considerations for $B^{-1}$ lead to
    \begin{eqaed}\label{eq:B_for_alpha+-1}
        B=\left(W(e^{3(\phi-c)})\right)^{-1}\mperiod
    \end{eqaed}
    \item For $\alpha\neq -1,-4$, one can integrate the second of \cref{eq:A_and_B_eq} in order to obtain an implicit definition of $B$:
    \begin{eqaed}\label{eq:B_for_generic_alpha}
        \frac{1}{1+\alpha}\log B -\frac{5+2\alpha}{(1+\alpha)(4+\alpha)}\log\left|B-\frac{1+\alpha}{4+\alpha}\right|=\phi-c\mperiod
    \end{eqaed}
\end{itemize}

We now focus on the relevant potentials.
Two values of $\alpha$, which are not our primary interests but are still of physical concern, are $\alpha=-\frac{5}{2}$ and $\alpha=-2$. 
The former, which reproduces a cosmological constant in the Einstein frame, simplifies \cref{eq:B_for_generic_alpha}, so that $B=e^{-\frac{3}{2} (\phi-c)}$, and the fake superpotential is a sum of two exponentials.
The latter, relevant for non-critical strings, simplifies \cref{eq:B_for_generic_alpha}, because when $-4<\alpha<1$ we do not need the absolute value, and yields $B^{-1}=\sqrt{1+2 e^{2(\phi-c)}}-1$.

For $\alpha=0$, which is significant for the $\text{SO}(16)\times\text{SO}(16)$ string and for vacuum energies, we have both a constant $B$ solution and the non-trivial one from \cref{eq:B_for_generic_alpha}.
The former, up to an overall sign in both $\mathcal{W}$ and $g$, is
\begin{eqaed}
    \mathcal{W}(\phi)&= \frac{1}{2}\sqrt{\frac{T}{8}}e^\phi\mcomma \\
    g(\phi)&= 5 \sqrt{\frac{T}{8}}e^\phi\mcomma
\end{eqaed}
and also applies to massive IIA supergravity, while its possible interpretation in the heterotic theory remains an open question~\cite{Raucci:2022jgw}.
The latter is more interesting from a non-supersymmetric perspective, but $B$ can be only expressed implicitly in \cref{eq:B_for_generic_alpha}. This leads to two possible behaviors, up to overall signs. One is a $B$ that decreases monotonically from infinity to $\frac{1}{4}$, corresponding to a $\mathcal{W}$ with a minimum. The other is a step-function-like $B$, interpolating between $0$ and $\frac{1}{4}$, which corresponds to a fake superpotential monotonically increasing from $0$ to infinity. We shall comment on their role in \cref{sec:other_solutions}.

For $\alpha=-1$, omitting the overall sign ambiguity, \cref{eq:B_for_alpha+-1} gives rise to
\begin{eqaed}\label[pluralequation]{eq:string_frame_fake_superpotential_gravity_scalar}
    \mathcal{W}(\phi) & = \frac{1}{6}\sqrt{\frac{T}{4}}e^{\frac{1}{2}\phi}\left[\sqrt{\frac{W(e^{3(\phi-c)})}{2}}+\sqrt{\frac{2}{W(e^{3(\phi-c)})}}\right]\mcomma \\
    g(\phi)& = \sqrt{\frac{T}{2}}e^{\frac{1}{2}\phi}\left[\sqrt{W(e^{3(\phi-c)})}+\sqrt{\frac{1}{W(e^{3(\phi-c)})}}\right]\mperiod
\end{eqaed}
The fake superpotential from \cref{eq:string_frame_fake_superpotential_gravity_scalar} is relevant for the two orientifold non-supersymmetric string models and has no critical points. In \cref{sec:vacuum_solutions}, we will use this explicit form to study the Dudas-Mourad solution. However, before proceeding, it is useful to reformulate our results in the Einstein frame, in view of the energy considerations of \cref{sec:energy}.

\subsection{Tadpole potential in the Einstein frame}

With a mild abuse of notation, we use Einstein-frame operators with the same names as the string-frame ones in \cref{eq:fake_susy_eq_variations}.
The metric and dilaton equations follow from the fake supersymmetry equations if
\begin{eqaed}\label[pluralequation]{eq:fake_susy_einst_frame_conditions}
    g+32\mathcal{W}' & = 0\mcomma \\
    2^9 (\mathcal{W}')^2 - 3^2 2^5 \mathcal{W}^2 & = V\mcomma \\
    2^{10} \mathcal{W}' \mathcal{W}'' - 3^2 2^6 \mathcal{W}\mathcal{W}' & = V'\mperiod
\end{eqaed}
The third of \cref{eq:fake_susy_einst_frame_conditions} follows from the second one, and $g$ is proportional to the derivative of $\mathcal{W}$. Therefore, one is left with a single differential equation for $\mathcal{W}$.
The tadpole-inspired exponential potential,
\begin{eqaed}
    V=T e^{\beta\phi}\mcomma
\end{eqaed}
induces different types of solutions depending on the value of $\beta$, reflecting the $\alpha$ dependence of \cref{eq:A_and_B_eq}. Let 
\begin{eqaed}\label{eq:definition_of_f}
    \mathcal{W}(\phi)=\frac{1}{12}\sqrt{\frac{T}{2}}e^{\frac{\beta}{2}\phi}\sinh{f(\phi)}\mperiod
\end{eqaed}
The differential equation reads
\begin{eqaed}\label{eq:Einstein_frame_fake_W_diffeq}
    16 (\cosh{f})^2 (f')^2+16\beta  f' \sinh{f} \cosh{f} + (4\beta^2-9)(\sinh{f})^2=9 \mcomma
\end{eqaed}
and
\begin{itemize}
    \item For $\beta^2>\frac{9}{4}$, there is a constant solution, matching the behavior in \cref{eq:constant_B}, with a fake superpotential $\sim e^{\frac{\beta}{2}\phi}$.
    \item The solution for $\beta=-\frac{3}{2}$ is the same as that for $\beta=\frac{3}{2}$ if one sends $\phi\to-\phi$, therefore we will only analyze the $\beta=\frac{3}{2}$ case.
    \item For $\beta=\frac{3}{2}$, \cref{eq:Einstein_frame_fake_W_diffeq} becomes
    \begin{eqaed}\label{eq:lambda_eq}
    \left[\frac{1}{3}(1+e^{-2f})2f'+1\right]\left[\frac{1}{3}(1+e^{2f})2f'-1\right]=0\mperiod
    \end{eqaed}
    The two solutions of \cref{eq:lambda_eq} are interchanged by $f\leftrightarrow -f$, which reflects a sign ambiguity in the definition of $f$, and hence of $\mathcal{W}$. Therefore, up to this sign,
    \begin{eqaed}\label{eq:lambda_sol}
    2f-e^{-2f}=-3\phi-3 c \Rightarrow 2f=W(e^{3(\phi-c)})-3(\phi-c)\mperiod
    \end{eqaed}
    This is the same case as $\alpha=-1$ in \cref{eq:B_for_alpha+-1}.
    \item For $\beta^2\neq\frac{9}{4}$, one can solve for $f'$ in \cref{eq:Einstein_frame_fake_W_diffeq} and, up to the usual $f\to-f$ sign ambiguity,
    \begin{eqaed}
        f'=-\frac{1}{2}\beta \tanh{f}-\frac{3}{4}\mcomma
    \end{eqaed}
    which leads to the implicit definition
    \begin{eqaed}\label{eq:f_for_generic_beta}
        -3f+2\beta\log\left|2\beta \sinh{f}+3\cosh{f}\right|=\left(\frac{9}{4}-\beta^2\right)(\phi-c)\mperiod
    \end{eqaed}
    This is the analogue of \cref{eq:B_for_generic_alpha}, and for $\beta^2<\frac{9}{4}$ one can remove the absolute value. In this subclass, one can find the cosmological constant and the non-critical string potentials, matching the results of the previous section.
\end{itemize}

The two exponents entering the non-supersymmetric string models are $\beta=\frac{3}{2}$ and $\frac{5}{2}$, respectively for the orientifold and heterotic strings. For the orientifold case, using properties of the $W$ function, the fake superpotential, up to the overall sign, takes the form
\begin{eqaed}\label{eq:einstein_frame_fake_W}
    \mathcal{W}(\phi) = \frac{1}{24}\sqrt{\frac{T}{2}}e^{\frac{3}{4}\phi}\left[\left(W(e^{3(\phi-c)})\right)^{-\frac{1}{2}} - \left(W(e^{3(\phi-c)})\right)^{\frac{1}{2}} \right]\mperiod
\end{eqaed}
It is a monotonically decreasing function with a non-vanishing first derivative.
For the heterotic case, we cannot be more explicit than in \cref{eq:f_for_generic_beta}, and, as in the string frame, there are two possible behaviors.

\subsection{A comment on fluxes}\label{sec:RR}

In the previous sections, we have restricted our analysis to the metric and the dilaton. The inclusion of form fields would be the natural and most interesting extension of the fake supersymmetry equations. In this section, we elaborate on the obstacles that apparently prevent us from obtaining a simple adjustment of \cref{eq:fake_susy_eq_variations}.
We shall work in the string frame, and for definiteness we shall mostly focus on the Sugimoto model. We shall briefly comment on the heterotic counterpart at the end of the section.
Let us anticipate that here, for simplicity, we shall refrain from introducing modifications to the kinetic terms, and our strategy will eventually fail.

Essentially, we must change the fake superpotential in such a way as to include the R-R forms in the equations of motion. In type I supergravity, in the democratic formalism, the gravitino and dilatino variations in the presence of non-vanishing R-R forms become, in the notation of~\cite{Tomasiello:2022dwe},
\begin{eqaed}\label[pluralequation]{eq:susy_type_I}
    \delta\psi_M&=D_{M}^{\mbox{\scriptsize susy}}  \varepsilon=\left(\nabla_M+\frac{1}{16}e^\phi F \Gamma_M\right)\varepsilon\mcomma \\
    \delta\lambda & =\mathcal{O}^{\mbox{\scriptsize susy}}  \varepsilon= \left(d\phi+\frac{1}{16}e^\phi\Gamma^M F\Gamma_M \right)\varepsilon\mcomma
\end{eqaed}
where $F=F_3+F_7$. 
The few terms present in \cref{eq:susy_type_I} provide the motivation for turning to type II theories for richer string compactifications, because they allow for more classical fluxes. From this perspective, the addition of a dilaton tadpole could be a blessing in disguise.

Unfortunately, the rigid structure of the supersymmetry equations is an obstacle for the type of extension that we have in mind. 
In fact, one can obtain the gravitational equations of motion from
\begin{eqaed}
    0& = \left[\Gamma^M\comm{D^{\mbox{\scriptsize susy}}_M}{D^{\mbox{\scriptsize susy}}_N}+\comm{\nabla_N}{\mathcal{O}^{\mbox{\scriptsize susy}}}-\frac{1}{16}e^\phi \mathcal{O}^{\mbox{\scriptsize susy}}F\Gamma_N\right]\varepsilon =\\
    & = \left[\frac{1}{2}(\mbox{Gravity EoM})_{MN}\Gamma^M +\frac{1}{16}e^\phi (dF) \Gamma_N \right]\varepsilon\mcomma
\end{eqaed}
with $F=F_3+F_7$ for type I, and the covariant derivative and the $\mathcal{O}$ operators defined in \cref{eq:susy_type_I}. 
The equations of motion are
\begin{eqaed}\label[pluralequation]{eq:equations_of_motion_type_I}
    R_{MN}+2\nabla_M\nabla_N\phi-\frac{1}{4}e^{2\phi}(F^2)_{MN} & = 0\mcomma \\
    R+4\nabla^2\phi-4(\d\phi)^2&=0\mcomma
\end{eqaed}
and the crucial contribution, which is the quadratic term in the R-R fields, originates from $\Gamma^M \comm{D^{\mbox{\scriptsize susy}}_M}{D^{\mbox{\scriptsize susy}}_N}$ and $e^\phi {O}^{\mbox{\scriptsize susy}}F\Gamma_N$.
Our hope would be to define two operators, as in \cref{eq:fake_susy_general}, which imply the equations of motion with the tadpole potential:
\begin{eqaed}\label[pluralequation]{eq:equations_of_motion_tadpole_and_fluxes}
    R_{MN}+2\nabla_M\nabla_N\phi+\frac{1}{2}e^{2\phi}\left(V+\frac{1}{2}V'\right) g_{MN}-\frac{1}{4}e^{2\phi}(F^2)_{MN} & = 0\mcomma \\
    R+4\nabla^2\phi-4(\d\phi)^2+\frac{1}{2}e^{2\phi}V'&=0\mperiod
\end{eqaed}
The new fake superpotential must then include the form fields, but working in full generality as in the flux-free case is much more difficult. Terms of the type $\comm{D_M}{D_N}$ and $e^{\phi}{O}F$, which would be relevant for the quadratic terms in the form fields, must produce the $e^{2\phi}$ factor in the first of \cref{eq:equations_of_motion_tadpole_and_fluxes}. The recovery of the operator relevant for the dilaton equation, which in the supersymmetric case is
\begin{eqaed}\label{eq:dilaton_eom_susy}
    (D^{\mbox{\scriptsize susy}}-\mathcal{O}^{\mbox{\scriptsize susy}}+\frac{1}{16}e^\phi F)(D^{\mbox{\scriptsize susy}}-\mathcal{O}^{\mbox{\scriptsize susy}})-(\nabla^M -2 \d^M\phi)D^{\mbox{\scriptsize susy}}_M\mcomma
\end{eqaed}
rests on the property that $D$ and $\mathcal{O}$ have the same $e^\phi$ prefactor in front of the R-R form fields.
These considerations prompt us to define, in the simplest possible way,
\begin{eqaed}\label[pluralequation]{eq:proposal_fake_susy_forms}
    D_{M}\varepsilon & =\left(\nabla_M + \mathcal{W}(\phi)\Gamma_M+\frac{1}{16}e^\phi F \Gamma_M\right)\varepsilon\mcomma \\
    \mathcal{O}\varepsilon&= \left(d\phi+g(\phi)+\frac{1}{16}e^\phi \Gamma^M F \Gamma_M\right)\varepsilon\mperiod
\end{eqaed}

Unfortunately, these expressions do not imply the equations of motion. One can verify indeed that
\begin{eqaed}\label{eq:fluxes_do_not_cancel}
    0&=\left[\Gamma^M\comm{D_M}{D_N}+\comm{\nabla_N+\mathcal{W}\Gamma_N}{\mathcal{O}}-\frac{1}{16}e^\phi \mathcal{O} F \Gamma_N+ (\mathcal{W}'-2\mathcal{W})\Gamma_N \mathcal{O}\right]\varepsilon=\\
    & =\bigg[ \frac{1}{2}(\mbox{Gravity EoM without } V(\phi))_{MN}\Gamma^M+(\propto\mbox{Bianchi identities}) \Gamma_N +\\
    & + (18\mathcal{W}^2+ \mathcal{W}'g - 2 \mathcal{W} g)\Gamma_N +(g'-2\mathcal{W}-8\mathcal{W}')\nabla_N\phi+\\
    & +\frac{1}{16}e^\phi \left((8\mathcal{W}-g)F \Gamma_N- \mathcal{W} \Gamma^M F \Gamma_N\Gamma_M+\mathcal{W}'\Gamma_N\Gamma^M F \Gamma_M)\right)\bigg] \varepsilon \mperiod
\end{eqaed}
The third line is the same contribution that appeared in \cref{eq:string_gravity_eom}, leading to the tadpole terms in the equations of motion if $\mathcal{W} $ and $g$ are defined by \cref{eq:the_three_conditions}.
However, the last line of \cref{eq:fluxes_do_not_cancel} does not vanish with these $\mathcal{W} $ and $g$. 
In fact, similar considerations hold for the dilaton equation, generalizing \cref{eq:dilaton_eom_susy}, for which one would need that
\begin{eqaed}\label{eq:fluxes_for_fake_dilaton_eom}
    (10 \mathcal{W}-g)F+ (g'-9\mathcal{W}'-\mathcal{W})\Gamma^M F \Gamma_M=0\mperiod
\end{eqaed}
Using the second of \cref{eq:the_three_conditions} in \cref{eq:fluxes_for_fake_dilaton_eom}, the vanishing of the last line of \cref{eq:fluxes_do_not_cancel} would imply $g=\mathcal{W}=0$, an indication that this strategy only works for the supersymmetric case, insofar as no other modifications are included beyond the tadpole potential itself.

There are in principle alternative routes, for instance increasing the number of spinors $\varepsilon_i$, but the author has been unable to overcome the structural problems that led to the ansatz in \cref{eq:proposal_fake_susy_forms}. One natural adjustment would be keeping the $e^\phi$ in front of the form fields, while leaving free numerical coefficients $a F\Gamma_M$, $b \Gamma_M F$, $c \Gamma^M F \Gamma_M$ and $d F$ in the fake supersymmetry equations. However, one would thus generate terms of the type $(F^2)_{MN}$ and $F^2 g_{MN}$ in the metric equations of motion, obtaining two equations for the four coefficients. Then, the requirement that $D-\mathcal{O}$ should not contain form fields would yield two more equations, fixing $a,b,c,d$ to the values that correspond to the supersymmetric case, thus recovering \cref{eq:proposal_fake_susy_forms}.

A similar fate emerges for the heterotic model, in which the $3$-form field belongs to the NS-NS sector and has no relative $e^{2\phi}$ factor. The simplest attempt,
\begin{eqaed}\label[pluralequation]{eq:heterotic_proposal_fake_susy}
    D_{M}\varepsilon & =\left(\nabla_M + \mathcal{W}(\phi)\Gamma_M+\frac{1}{4}H_M\right)\varepsilon\mcomma \\
    \mathcal{O}\varepsilon&= \left(d\phi+g(\phi)+\frac{1}{2} H \right)\varepsilon\mcomma
\end{eqaed}
would be consistent with the equations of motion only for $\mathcal{W}=0$, which rules out the scalar potential.

These considerations suggest that, if a first-order formalism including form fields exists, it necessarily departs from the simple form of the supersymmetric variations. We suspect that further progress will require that one reconsider the structure of the kinetic terms, allowing in them more general scalar contributions. We plan to return to this point soon.

\section{Vacuum solutions}\label{sec:vacuum_solutions}

One of the most basic questions that one can ask in string compactifications is whether there exists a vacuum solution, a set of relatively symmetric field profiles describing an empty spacetime. This usually translates into a vanishing stress energy tensor, but for non-supersymmetric strings we know that the tadpole potential cannot disappear. In some sense, it is a cosmological constant with a non-zero coupling to the dilaton, so that a vacuum solution should naturally involve a non-trivial profile for the dilaton. However, the form of the potential and the absence of minima make the analysis more involved, and at the same time more interesting, for instance possibly leading to the need for time dependence.

Before proceeding, let us review how vacuum solutions arise in the supersymmetric case. One usually considers a metric ansatz with a four-dimensional extended maximally symmetric space 
\begin{eqaed}
    ds^2=e^{2A}ds_4^2+ ds_6^2\mcomma
\end{eqaed}
with the warping function $A$ and the dilaton $\phi$ depending on the coordinates of the six-dimensional manifold. If one is interested in vacua with unbroken supersymmetry, the first step is to realize how spinors decompose. The space of ten-dimensional spinors is a tensor product between the spaces of four and six-dimensional spinors, therefore one can consider $\Gamma$ matrices of the form
\begin{eqaed}
    \Gamma_\mu=e^A \gamma_\mu \otimes 1\mcomma \qquad \Gamma_i=\gamma^{(4)}\otimes \gamma_i\mcomma
\end{eqaed}
with $\gamma^{(4)}$ the chirality matrix in four dimensions, along with the spinor decomposition
\begin{eqaed}\label{eq:susy_spinor_ansatz}
    \varepsilon=\zeta_+ \otimes \eta_+ + \zeta_- \otimes \eta_-\mperiod
\end{eqaed}
The notation for the chiral spinors $\zeta $ and $\eta$ is to be understood schematically, because there might be additional indices representing the amount of supersymmetry, or $\varepsilon$ might even be defined as a sum over a set of $\zeta_I$ and $\eta_J$. The Majorana property demands that $(\zeta_\pm)^c=\zeta_{\mp}$, and the same for $\eta$. Then, because of the isometries of the vacuum solution, $\zeta$ must live in the space of four-dimensional Killing spinors, and $\eta$ must only depend on the internal coordinates.
Therefore, defining $\nabla_\mu \zeta_\pm =\frac{\mu}{2}\gamma_\mu \zeta_\mp$ and assuming $\d_i\zeta_\pm=0$, unbroken supersymmetry requires
\begin{eqaed}\label[pluralequation]{eq:gravitino_dilatino_in_susy}
    \nabla_M\varepsilon & = 0\mcomma \\
    d\phi \varepsilon&=0 \mperiod
\end{eqaed}
The first of \cref{eq:gravitino_dilatino_in_susy}, using the spinor ansatz in \cref{eq:susy_spinor_ansatz}, leads to
\begin{eqaed}
    \nabla_i \eta_+ &=0\mcomma \\
    \mu e^{-A}\eta_+ - dA \cdot \eta_-&=0\mperiod
\end{eqaed}
Then, since $\eta_+$ and $\gamma^i \eta_-$ are independent spinors, this is equivalent to 
\begin{eqaed}
    \mu=0\mcomma \qquad dA=0\mcomma \qquad  \nabla_i \eta_+ =0\mperiod
\end{eqaed}
The preceding steps recover the well-known result that supersymmetric vacuum solutions are unwarped four-dimensional Minkowski spaces with internal six-dimensional manifolds with reduced holonomy~\cite{Candelas:1985en}. The dilatino variation in \cref{eq:gravitino_dilatino_in_susy}  requires a constant dilaton, which concludes the analysis.

Given the results of \cref{sec:fake_susy}, one might want to replicate the reasoning for the non-supersymmetric strings employing our first-order formalism. However, there is a problematic challenge that is not unexpected, due to the structure of the fake susy equations in \cref{eq:fake_susy_eq_variations}. In fact, in both equations, the new terms $\mathcal{W}\Gamma_M$ and $g$ carry a number of gamma matrices with opposite parity with respect to the potential-free case, therefore inducing too many conditions on the spinor. Explicitly, assuming the same ansatz as before, in \cref{eq:susy_spinor_ansatz}, \cref{eq:fake_susy_eq_variations} yield
\begin{eqaed}
    \mu e^{-A}\eta_- + dA \eta_+ +2\mathcal{W} \eta_+&=0\mcomma \\
    \nabla_i \eta_+ + \mathcal{W} \gamma_i \eta_+ & = 0\mcomma \\
    (d\phi+g)\eta_+ & = 0\mperiod
\end{eqaed}
In the third equation, $\eta$ and $\gamma_m\eta$ are independent, and the only possible solution requires $d\phi=0$, which is however inconsistent with a non-zero $g$.
Similarly, the first equation requires $\mu=0$ and $A=0$, which is again incompatible with the presence of $\mathcal{W}$. 
We then discover that the first-order formalism that we introduced does not contain any vacuum solution analogous to the Minkowski $\times$ CY vacua of supersymmetric string theory.

There are various approaches that one might want to follow, after the failure of the simplest procedure. One possibility would be to modify the spinor ansatz in \cref{eq:susy_spinor_ansatz}, making the analysis less natural, or even to work in the four-dimensional low-energy effective theory with non-trivial kinetic terms for the compactification scalars. A more basic possibility would be to invoke the necessity of fluxes, abandoning the idea of susy-like vacuum solutions from the first-order formalism.
In this section, we shall instead explore a setting that evades the above issues: codimension-one vacua.

\subsection{Codimension one}

Consider a codimension-one ansatz
\begin{eqaed}
    ds^2&=e^{2A(y)}dx^2 + dy^2\mcomma\\
    \phi & = \phi(y)\mcomma
\end{eqaed}
with a ten-dimensional spinor $\varepsilon(y)$ that depends on the coordinate $y$. 
This ansatz seems more likely to produce a non-trivial solution of the first-order equations. In fact, the literature on fake supersymmetry is primarily devoted to the study of domain walls, with asymptotically flat or $AdS$ spacetimes. 
See for instance~\cite{Freedman:2003ax}, where spinors are asymptotically Killing spinors and the fake superpotential becomes a constant. 
In our settings, we do not have a clear understanding of the asymptotic boundary conditions, which will become a problem in \cref{sec:energy}, but for the moment we can still apply the first-order equations to the codimension-one ansatz, at least far from the possible locations of domain walls.

We can safely consider \cref{eq:fake_susy_eq_variations} in either the Einstein or string frames, because our choice of notation allows this ambiguity. Making use of the codimension-one ansatz for metric, dilaton and spinor $\varepsilon$, the fake supersymmetry equations lead to 
\begin{eqaed}\label[pluralequation]{eq:codim_1_fake_W}
    \phi'(y) & = \pm g(\phi)\mcomma\\
    \d_y \varepsilon & = \pm \mathcal{W} \varepsilon\mcomma \\
    A'(y) & = \pm 2\mathcal{W} \mperiod
\end{eqaed}
Then, one can write
\begin{eqaed}
    \varepsilon=e^{\frac{1}{2}A(y)}\varepsilon_0\mcomma    
\end{eqaed}
with $\varepsilon_0$ a constant ten-dimensional spinor.
These considerations are valid independently of the explicit form of the scalar potential.

\subsection{The Dudas-Mourad solution}\label{sec:Dudas-Mourad}

The Dudas-Mourad solution~\cite{Dudas:2000ff} is a codimension-one vacuum for the tachyon-free non-supersymmetric string theories in ten dimensions. It is the natural arena for the formalism that we have developed, and being perturbatively stable~\cite{Basile:2018irz}, it is also an appropriate example for the stability questions that we mentioned in the Introduction, and that we shall explore further in \cref{sec:energy}.
We focus on the tadpole potential for the orientifold models, both for definiteness and because this is the case in which we have an explicit fake superpotential.
The first comforting result comes from the differential equations for the fake superpotential, because both \cref{eq:the_three_conditions} in the string frame and \cref{eq:fake_susy_einst_frame_conditions} in the Einstein frame become the Dudas-Mourad equations of motion for $A$ and $\phi$ in the gauge $B=0$, as expected from the general construction of \cref{sec:fake_susy}.

In order to explicitly compare the codimension-one vacuum of~\cite{Dudas:2000ff} with our fake superpotential, we must turn to the appropriate gauge, $B=-\frac{1}{2}\phi$ in the string frame or $B=-\frac{3}{4}\phi$ in the Einstein frame. We choose to work in the string frame, in which the Dudas-Mourad solution reads
\begin{eqaed}\label[pluralequation]{eq:Dudas_Mourad}
    ds_s^2&=\left(\sqrt{\frac{T}{2} }y\right)^{\frac{4}{9}}e^{\frac{1}{2}\phi_0}e^{\frac{T}{8}y^2}dx^2 + \left(\sqrt{\frac{T}{2}}y\right)^{-\frac{2}{3}}e^{-\phi_0}e^{-\frac{3T}{8}y^2}dy^2\mcomma\\
    \phi&=\phi_0+\frac{3T}{8}y^2 +\frac{2}{3}\log\left(\sqrt{\frac{T}{2}}y\right)\mperiod
\end{eqaed}
The second of \cref{eq:fake_susy_eq_variations}, using the expression for $g(\phi)$ in \cref{eq:string_frame_fake_superpotential_gravity_scalar}, is
\begin{eqaed}\label{eq:DM_condition1}
    e^{\frac{1}{2}\phi}\phi' \gamma_y\varepsilon=\mp\sqrt{\frac{T}{2}}e^{\frac{1}{2}\phi}\left[\sqrt{W(e^{3(\phi-c)})}+\sqrt{\frac{1}{W(e^{3(\phi-c)})}}\right]\varepsilon\mperiod
\end{eqaed}
The explicit dilaton profile in \cref{eq:Dudas_Mourad}, along with some properties of the $W$ function, determines the integration constant $c$ in terms of $\phi_0$ as $c=\phi_0+\frac{2}{3}\log\frac{2}{3}$. Then, \cref{eq:DM_condition1} is equivalent to $\gamma_y\varepsilon=\mp\varepsilon$, with $\gamma_y$ the gamma matrix in flat space.
The fake superpotential and the $g$ function, written in terms of $\phi_0$, read
\begin{eqaed}\label[pluralequation]{eq:DM_string_frame_fake_superpotential}
    \mathcal{W}(\phi) & = \pm\frac{1}{12}\sqrt{T}e^{\frac{1}{2}\phi}\left[\sqrt{\frac{W(\frac{9}{4}e^{3(\phi-\phi_0)})}{2}}+\sqrt{\frac{2}{W(\frac{9}{4}e^{3(\phi-\phi_0)})}}\right]\mcomma \\
    g(\phi)& = \pm\sqrt{T}e^{\frac{1}{2}\phi}\left[\sqrt{\frac{W(\frac{9}{4}e^{3(\phi-\phi_0)})}{2}}+\sqrt{\frac{1}{2 W(\frac{9}{4}e^{3(\phi-\phi_0)})}}\right]\mperiod
\end{eqaed}
We now turn to the first of \cref{eq:fake_susy_eq_variations} with $M=\mu$, which is
\begin{eqaed}\label{eq:DM_condition2}
    \frac{1}{2}e^{\frac{1}{2}\phi}A' \gamma_y\varepsilon=\mp\frac{1}{12}\sqrt{T}e^{\frac{1}{2}\phi}\left[\sqrt{\frac{W(\frac{9}{4}e^{3(\phi-\phi_0)})}{2}}+\sqrt{\frac{2}{W(\frac{9}{4}e^{3(\phi-\phi_0)})}}\right]\varepsilon\mcomma
\end{eqaed}
and is also equivalent to $\gamma_y\varepsilon=\mp\varepsilon$. On the other hand, the $M=y$ equation contains the $y$-dependence of the spinor, and produces
\begin{eqaed}\label{eq:DM_condition3}
    \d_y \varepsilon=\left(\frac{T}{16}y+\frac{1}{9y}\right)\varepsilon\mperiod
\end{eqaed}
To summarize, our first-order formalism captures the Dudas-Mourad solution with \cref{eq:DM_string_frame_fake_superpotential}, and corresponds to a string-frame spinor 
\begin{eqaed}\label[pluralequation]{eq:DM_String_frame_spinor}
    \varepsilon_S&=e^{\frac{T}{32}y^2}\left(\sqrt{\frac{T}{2}}y\right)^{\frac{1}{9}}\varepsilon_0\mcomma\\
    \gamma_y\varepsilon_0&=\mp\varepsilon_0\mcomma
\end{eqaed}
with the two signs corresponding to the overall sign ambiguity of $\mathcal{W}$.
For completeness, and in view of \cref{sec:energy}, the Einstein-frame spinor in the gauge and the variables of~\cite{Dudas:2000ff} is
\begin{eqaed}\label[pluralequation]{eq:DM_Einstein_frame_spinor}
    \varepsilon_E&=e^{-\frac{T}{64}y^2}\left(\sqrt{\frac{T}{2}}y\right)^{\frac{1}{36}}\varepsilon_0\mcomma\\
    \gamma_y\varepsilon_0&=\mp\varepsilon_0\mperiod
\end{eqaed}

\subsection{Other solutions}\label{sec:other_solutions}

In \cref{sec:fake_susy}, we studied a generic exponential potential. The codimension-one equations of motion for these systems were recently analyzed in~\cite{Basile:2022ypo}, and we might ask whether those solutions are captured by the fake superpotentials that we obtained. The answer is positive, and in \cref{sec:some_comments} we shall mention the reason why this is not surprising. For the time being, let us find the explicit matching.

A simple case is the superpotential corresponding to \cref{eq:constant_B}, which provides the supercritical solutions of~\cite{Basile:2022ypo} with linear $A$ and $\phi$. In fact, the gauge used in~\cite{Basile:2022ypo} is such that the function $f(y)$ of~\cite{Basile:2022ypo} is the same as the $f(\phi(y))$ that we use in \cref{eq:definition_of_f}, which partly explains our notation in this paper. The differential equation~(\ref{eq:Einstein_frame_fake_W_diffeq}) matches the differential equation for $f(y)$ in~\cite{Basile:2022ypo}.

The correspondence survives in the generic case because the gauge $B=-\frac{\beta}{2}\phi$ in the Einstein frame generates a $(D_\mu+\mathcal{W}\Gamma_\mu)\varepsilon=0$ equation, which, using \cref{eq:definition_of_f}, reduces to 
\begin{eqaed}
A'(y)\propto \sinh{f(\phi(y))}\mperiod
\end{eqaed}
Using the explicit results in~\cite{Basile:2022ypo}, one can then verify that the matching is complete, provided that the spinors are
\begin{eqaed}
    \varepsilon(y)=e^{\frac{A(y)}{2}}\varepsilon_0\mcomma
\end{eqaed}
in complete agreement with \cref{eq:codim_1_fake_W}.
In particular, the two non-trivial fake superpotentials for the heterotic theory that we derived in \cref{sec:tadpole_string_frame} correspond to the two independent types of codimension-one solutions of~\cite{Dudas:2000ff}, and the monotonic $\mathcal{W}$ represents the less interesting unbounded solution. 

A similar analysis could be accomplished with the scalar potentials of~\cite{Pelliconi:2021eak}, computing the fake superpotentials relevant in those cases, and we leave this investigation for future work.

Note that even when $T=0$ there exists a non-vanishing $\mathcal{W}$, which in the Einstein frame is $\mathcal{W}_0 e^{\pm \frac{3}{4}\phi}$. This reproduces the uncharged codimension-one solutions of~\cite{Raucci:2022jgw}.

\subsection{Some comments on the formalism}\label{sec:some_comments}

We have found an exact matching between codimension-one solutions obtained from the equations of motion and from the fake supersymmetry approach. 
This is not coincidental, and in fact in~\cite{Townsend:2007aw, Trigiante:2012eb} the authors found that, for a class of gravity solutions that depend on a single coordinate, one can trade the second-order classical equations of motion for the same number of first-order equations, together with a non-linear partial differential equation. 
This non-linear differential equation connects, in fact, the scalar potential and the corresponding fake superpotential.
One can link the class of gravitational systems for which this procedure is possible to the vanishing of a conserved charge in the Hamilton-Jacobi formulation of the associated one-dimensional mechanical system.\footnote{See~\cite{Trigiante:2012eb} for a detailed explanation of how this property is related to the factorization of Hamilton's principal function.} 

The present setting is more general, because the first-order equations of~\cite{Trigiante:2012eb} need not assume the form of \cref{eq:fake_susy_eq_variations}. In fact, in the fake supersymmetry literature one usually handles the differential equations for the fields entering the metric ansatz, while our first-order formalism is supposed to work with any metric ansatz. 
This is ultimately the reason behind the difficulty that we encountered with form fields, but if we managed to circumvent the obstacles of \cref{sec:RR}, it might open a new window into the non-supersymmetric string landscape.

\section{Energy and stability}\label{sec:energy}

As we explained in the Introduction, our investigation attempts to replace supersymmetry with other criteria granting the dynamical stability of string vacua.

Assessing the complete stability would require a proper control of quantum effects, which is difficult to attain in general, and in particular in a theory of gravity, where our physical intuition, however, can capture some details of the (semi)classical regime.
In regions of field space where perturbative string theory is a good description of quantum gravity, the absence of tachyons in the spectrum characterizes stable compactifications ending up in Minkowski space. The inclusion of non-perturbative effects is usually possible only relying on non-renormalization properties of supersymmetric vacua, which can usually upgrade some classical results to the quantum regime.
The first-order formalism that we developed in \cref{sec:fake_susy}, being only a formal tool, cannot replace supersymmetry in its physical content. However, it can still play a role for stability, because it can effectively mimic supersymmetry in the definition of a positive-definite quantity associated to vacua, which one might interpret as an energy.
This procedure is under control for asymptotically flat or $AdS$ vacua, but necessarily relies on non-trivial assumptions in all other cases.
The absence of a general definition of energy for a string vacuum\footnote{A string vacuum is a 2D SCFT, defining the background in which perturbative string theory lives. It would then be interesting to define a notion of energy directly from the worldsheet, which should match the known spacetime energies for the cases under control.} motivates us to explore the one that we can define with the operators $D_M$ and $\mathcal{O}$.

The absence of a global group of isometries, in a generic curved spacetime, prevents one from identifying the energy as the charge associated with time translations.
In fact, the type of energy that we shall use in this section for the dilaton-gravity setup with a tadpole potential is inspired by the Witten-Nester formalism.
The main idea is that, in a supersymmetric theory, a preserved real supercharge $Q$ is such that schematically $\left\langle\acomm{Q}{Q}\right\rangle\geq0$, and a supersymmetric state saturates the bound.
In~\cite{Witten:1981mf}, Witten used this spinor definition of energy, motivated by the above supersymmetry argument, addressing the stability of pure gravity.
Indeed, the various spinors appearing in the formalism can be taken as auxiliary objects, unrelated to any physical symmetries, and the only requirement is their existence.

Let us briefly review the general scheme in the case of four-dimensional asymptotically flat spacetimes, following the conventions of~\cite{Giri:2021eob}, in which the authors proved the stability of ten-dimensional supergravity vacua in analogous ways.
The energy is
\begin{eqaed}
    I(\varepsilon)\sim\left\langle\acomm{Q}{Q}\right\rangle\sim\int_{\d\Sigma}\star E_2\mcomma
\end{eqaed}
with a two-form
\begin{eqaed}
    E_2=-\frac{1}{2}\bar{\varepsilon} \, \gamma_{\mu\nu\rho}\nabla^\rho \varepsilon  \, dx^\mu \wedge dx^\nu\mcomma
\end{eqaed}
where 
\begin{itemize}
    \item $\varepsilon$ is a commuting Majorana spinor, which becomes a constant $\varepsilon_0$ at infinity, up to $r^{-1}$ terms, where $r$ is the radial distance.
    \item $\d\Sigma$ is the asymptotic boundary of a spatial slice. One can understand the surface $\Sigma$ as an initial-value surface.
\end{itemize} 
This is a consistent notion of energy, because
\begin{eqaed}
    I(\varepsilon)=I(\varepsilon_0)=-\bar{\varepsilon}_0\gamma^\mu \varepsilon_0 P_\mu\mcomma
\end{eqaed}
with $P$ the ADM four-momentum. 
In order to verify that $I(\varepsilon)$ is positive definite, one can apply Stokes' theorem, assuming the appropriate levels of smoothness, and then
\begin{eqaed}
    I(\varepsilon)=\int_\Sigma d\star E_2=\int_\Sigma \left(\nabla^\nu \bar\varepsilon \gamma_{\mu\nu\rho}D^\rho\varepsilon+\frac{1}{2}(R_{\mu\nu}-\frac{1}{2}R g_{\mu\nu})\bar{\varepsilon}\gamma^\nu \varepsilon\right)n^\mu\mcomma
\end{eqaed}
where $n^\mu$ is a time-like unit vector orthogonal to $\Sigma$. Consequently, the contribution proportional to the equations of motion vanishes and, in a frame where $n^\mu$ is directed toward the time direction, 
\begin{eqaed}
    I(\varepsilon)=\int_\Sigma (\nabla^i\varepsilon)^\dagger (\nabla_i \varepsilon)-\int_\Sigma |\gamma^i\nabla_i \varepsilon|^2\mcomma
\end{eqaed}
where $i$ denotes a spatial index. Positivity follows if 
\begin{eqaed}
    \gamma^i \nabla_i \varepsilon=0\mcomma
\end{eqaed}
which is known as the Witten condition.
In a supergravity theory, this condition would be equivalent to the choice of a transverse gauge for the gravitino, $\gamma^i\psi_i=0$.
Therefore, $I\geq0$, and $I=0$ only when $\nabla_i\varepsilon=0$. Then, the arbitrariness of $\Sigma$ implies that solutions with zero energy are supersymmetric vacua $\delta\psi_\mu=0$, concluding the proof of the positive energy theorem for Minkowski spacetimes.

The results of \cref{sec:fake_susy} prompt us to investigate a similar definition with a non-vanishing dilaton tadpole, starting from our first-order formalism. In fact, the same inspiration was the central topic of~\cite{Boucher:1984yx}, which elaborated on the observation that the Witten-Nester construction does not rely on supersymmetry, so that the spinors can be mere auxiliary variables.
Therefore, having in mind the effective theories for the non-supersymmetric strings with only gravity and the dilaton, in the Einstein frame, we can define 
\begin{eqaed}\label{eq:NW_2_form}
   E^{MN}=-\bar{\varepsilon}\Gamma^{MNP}D_P \varepsilon\mperiod
\end{eqaed}
This is the equivalent of the Witten-Nester two-form, with $D_P$ the modified derivative including the fake superpotential of \cref{eq:fake_susy_eq_variations}.

The usual procedure that we reviewed would transform a codimension-two (hyper)surface integral into a codimension-one boundary integral, using Stokes' theorem. However, this is consistent only when the codimension-one surface is non-singular, which is not the case for the vacua of \cref{sec:vacuum_solutions} that we want to analyze. One can include horizons, and even time-like singularities corresponding to BPS sources, but for the type of singularities that we encounter in the non-supersymmetric strings, we are not allowed to use Stokes' theorem. Therefore, we propose to write the energy directly in terms of an integral on a codimension-one surface, as 
\begin{eqaed}\label{eq:fake_energy}
    I(\varepsilon)=\int_\Sigma \nabla_N E^{MN} d\Sigma_M\mperiod
\end{eqaed}
The integral over $\Sigma$, in the cases of interest, still encloses singularities at the boundary, and this will inevitably generate some problems of interpretation. Postponing this discussion to \cref{sec:stability_DM}, from \cref{eq:NW_2_form} and \cref{eq:fake_susy_eq_variations} one can see that
\begin{eqaed}\label{eq:divergence_two_form}
    \nabla_M E^{MN}  = \overbar{D_M\varepsilon}\Gamma^{MPN}D_P\varepsilon+\frac{1}{2}\bar{\varepsilon}\left({\mbox{Gravity EoM}}\right)^{MN}\Gamma_M\varepsilon-\frac{1}{8}\overbar{\mathcal{O}\varepsilon}\Gamma^N \mathcal{O}\varepsilon\mperiod
\end{eqaed}

In a frame where the surface is purely spacelike, the on-shell value of energy in \cref{eq:fake_energy} becomes
\begin{eqaed}\label{eq:energy_from_NW}
    I(\varepsilon) = \int_\Sigma -\overbar{D_m\varepsilon}\Gamma^{0mp} D_p\varepsilon+\frac{1}{8}\overbar{\mathcal{O}\varepsilon}\Gamma^0 \mathcal{O}\varepsilon\geq \int_\Sigma -\left(D_m\varepsilon\right)^\dagger \Gamma^{m}\Gamma^{p}\left(D_p\varepsilon\right)\mperiod
\end{eqaed}
If we can impose the Witten condition with the new operator $D_M$,
\begin{eqaed}\label{eq:Witten_condition}
    \Gamma^m D_m\varepsilon=0\mcomma
\end{eqaed}
\cref{eq:energy_from_NW} provides a positive-definite quantity, with a natural interpretation as an energy. We shall not address here the technical questions related to the existence, in general, of such spinors. Note that $I(\varepsilon)$ vanishes if and only if the fake supersymmetry equations are satisfied, and we have seen that this is the case for the Dudas-Mourad vacuum.

\subsection{Implications for codimension one}\label{sec:stability_DM}

It is tempting to regard the analysis of the previous section as a proof that the Dudas-Mourad solution of \cref{eq:Dudas_Mourad} is a stable vacuum of non-supersymmetric string theories. We believe, however, that this implication would be too naive, for reasons related to the boundary conditions.
In fact, even assuming that \cref{eq:fake_energy} serves as a definition of energy, we can only infer that zero-energy configurations are stable under decay processes preserving the boundary conditions. In the Dudas-Mourad case, these contain the singular endpoints of the compact interval. Accordingly, we can trust the stability claim only if the timelike codimension-one singularities are physical and represent fixed boundaries. 
With a change of perspective, one can reformulate the previous statement linking the stability of the Dudas-Mourad solution to properties of the boundary.
This agrees with similar claims in the literature, and we think that our argument strengthens this physical intuition.

As a comment on the pivotal role of boundary conditions, we mention that even in some less puzzling setups there are no implications for stability, despite the presence of fake supersymmetry equations. In~\cite{Freedman:2003ax}, the authors considered a cubic fake superpotential, with the corresponding scalar potential yielding an $AdS$ vacuum. They showed that, if the $AdS$ vacuum is not an extremum of the fake superpotential as well, tachyonic instabilities below the B-F bound can arise. In this sense, the first-order formalism itself is not powerful enough unless the boundary conditions are under control.

Before concluding, we note that interpreting the Dudas-Mourad vacuum as a spontaneous compactification on an interval, and studying the nine-dimensional resulting effective field theory, as in~\cite{Basile:2022ypo}, our proposal for the energy would become the usual Witten-Nester energy for asymptotically flat spacetimes.
In fact, the remarkable property of the spinor $\varepsilon(y)$ in \cref{eq:DM_Einstein_frame_spinor} is that it is smooth, even though the gravity solution is singular. Then, integrating out the $y$-direction would lead to the nine-dimensional Witten-Nester energy, taking $\varepsilon_0$ as $\varepsilon_0(x)$.
This proves, for instance, that if a bubble of nothing exists for this class of vacua, it necessarily involves the $y$-interval in a non-trivial fashion.

\section{Conclusions}

This work is an attempt to setup a novel approach to the analysis of non\hyp{}supersymmetric string vacua, based on first-order susy-like equations. 
The basic formalism is not new, but we have applied it in a different way to string models with tadpole potentials.

Codimension-one vacua are the natural arena for the first-order formalism, because of the restriction to gravity and the dilaton. Unfortunately, this work did not generate so far new classes of vacua, for the reasons explained in \cref{sec:some_comments}.
Nonetheless, it led us to a peculiar definition of energy, which appears to encode useful information on vacuum stability in this context.
At any rate, our study of exponential potentials in \cref{sec:fake_susy}, and the explicit solution for the critical case in \cref{eq:string_frame_fake_superpotential_gravity_scalar}, are a step toward new ways to engineer non-supersymmetric vacua without relying on the second-order equations of motion.

Adding fluxes to the fake supersymmetry equations would clearly represent a considerable advance. Our work in \cref{sec:RR} is not the most general approach, as we have stressed, and generalizations could involve a different spinor ansatz or scalar-dependent extensions of kinetic terms.
We plan to return to this issue in future work.
A similar approach might be of interest in an effective field theory with more than one scalar, even in presence of a non-trivial scalar-field geometry.
A different question is what types of higher-derivative corrections can be included in this formalism, in view of $\alpha'$ corrections to the string equations of motion.
This could well impact the stability side, as is the case in other contexts.

As a final comment, we note that the flux solutions of~\cite{Mourad:2016xbk} are either perturbatively or non-perturbatively unstable~\cite{Basile:2018irz, Antonelli:2019nar}. Therefore, if one could generalize \cref{eq:fake_susy_eq_variations} with the inclusion of R-R and NS-NS forms, the first-order formalism should exclude these vacua. The connection between this prediction and the general procedure described in~\cite{Giri:2021eob} for addressing the stability of non-supersymmetric vacua deserves further scrutiny.

\section*{Acknowledgements}

We are grateful to A.~Tomasiello for comments and guidance in various parts of this work. We also thank I.~Basile, M.~Graña, A.~Herráez, N.~Kovensky, G.~Pimentel and A.~Sagnotti for stimulating discussions. We would also like to thank INFN-Milano Bicocca and IPhT-Saclay for the kind hospitality while this work was in progress.
This work was supported in part by Scuola Normale, by INFN (IS GSS-Pi) and by the MIUR-PRIN contract 2017CC72MK\_003.

\bibliographystyle{utphys}
\bibliography{mybib}

\end{document}